\documentclass[12pt]{JHEP3} 
\usepackage{epsfig,multicol,bbm}
\usepackage{latexsym}
\usepackage{amssymb}
\usepackage{amsfonts}
\usepackage{amsmath}
\usepackage{bm}
\usepackage{times}
\usepackage{graphicx}
\usepackage{axodraw}

\title{Confusing the extragalactic neutrino flux limit
with a neutrino propagation  limit}

\author{Juan Barranco\\
Instituto de Astronom\'{\i}a, Universidad Nacional Autonoma de
M\'exico, Mexico, DF 04510, Mexico}

\author{Omar G. Miranda\\
Departamento de F\'{\i}sica, Centro de Investigaci{\'o}n y de Estudios
Avanzados del IPN, Apdo. Postal 14-740 07000 M\'exico, D.F., Mexico}
\author{Celio A. Moura\\

Centro de Ci\^encias Naturais e Humanas, Universidade Federal do ABC,
Rua Santa Ad\'elia, 166, 09210-170 Santo Andr\'e, SP, Brazil}

\author{Timur I. Rashba\thanks{On leave from
IZMIRAN, Institute of Terrestrial Magnetism, Ionosphere and Radio Wave
Propagation of the Russian Academy of Sciences, 142190, Troitsk,
Moscow region, Russia}\\
Max-Planck-Institute for Solar System Research, Katlenburg-Lindau,
37191, Germany}

\author{Fernando Rossi-Torres\\
Instituto de F\'{\i}sica Te\'orica, Universidade Estadual Paulista,
Rua Dr. Bento Teobaldo Ferraz, 271 - Bl. II, 01140-070, S\~ao Paulo,
SP, Brazil}

\date{\today} 
\abstract{We study the possible suppression of the
extragalactic neutrino flux due to a nonstandard interaction during
its propagation. In particular, we study neutrino interaction with
an ultra-light scalar field dark matter. It is shown that the
extragalactic neutrino flux may be suppressed by such an
interaction, leading to a new mechanism to reduce the ultra-high
energy neutrino flux. We study both the cases of non-self-conjugate
as well as self-conjugate dark matter. In the first case, the
suppression is independent of the neutrino and dark matter
masses. We conclude that care must be taken when explaining limits
on the neutrino flux through source acceleration mechanisms only,
since there could be other mechanisms for the reduction of the
neutrino flux.}


\begin{document}

\section{Introduction}

It is commonly believed that Ultra-high Energy (UHE) neutrinos 
should arrive at the Earth coming from very distant sources like 
Active Galactic Nuclei (AGN) and
Gamma Ray Bursts (GRB) following straight trajectories. 
Neutrinos interact only through weak interactions. 
They are not charged and if they have a nonzero magnetic 
moment it must be very small. Hence, there would be no 
interaction preventing them to travel
cosmological distances~\cite{Halzen:2002pg}. 
The efforts to improve sensitivity to the UHE
neutrino flux may test the existence of the Berezinsky-Zatsepin
(BZ) neutrinos~\cite{Berezinsky}. Those neutrinos are generated through
the same process that predicts the Greisen-Zatsepin-Kuzmin (GZK)
cutoff~\cite{Greisen:1966jv} of Cosmic Ray (CR) flux, which has been
observed by HiRes~\cite{hires} and the Auger
Observatory~\cite{Abraham:2008ru}. The most energetic CRs, most probably
composed by protons and nuclei, may have an origin at relatively close
sources, of the order of 100~Mpc for protons and even less for
nuclei~\cite{Hooper:2006tn}. In fact, UHE CRs are most probably
arriving from nearby AGNs~\cite{Cronin:2007zz}. 
But we observe that despite all the efforts to detect high energy
extragalactic neutrinos, i.e., neutrinos with energies higher than
$10^{3}$~TeV, more and more restricted limits on their
flux have been reported by several experiments due to the non
observation of these neutrinos~\cite{Ackermann:2005sb,Desai:2007ra,
Abraham:2007rj,BlanchBigas:2009zz,Abbasi:2011qc}.

In addition, based on astronomical and cosmological observations, it
is difficult to deny the existence of Dark Matter (DM) and Dark Energy
(DE). Cosmological observations of clusters of galaxies indicate that
the density fraction of DM is $\Omega_{\rm DM} \approx 0.227 \pm
0.014$~\cite{wmap}. The two most popular solution to the DM problem
are {\it i)} a slight modification on the Newtonian dynamics
\cite{milgron} and {\it ii)} relic particles~\cite{Bertone:2004pz}. A
DM relic particle candidate must be non-relativistic, must be stable
on the cosmological time scale, and must interact weakly with other
particles. Some of the possible candidates are:
WIMPS~\cite{Jungman:1995df},
axions~\cite{Peccei:1977ur,Peccei:1977hh,Mikheev:2008zz,Andrianov:2009kj,dfsz},
MeV-scalar fields~\cite{Boehm:2004,Boehm:2003hm}, technicolor
candidates ~\cite{Foadi:2008qv,techndm1,techndm2}, and ultra-light
scalar fields~\cite{Matos:2008ag,Arbey:2001qi,Amendola:2005ad,sflocal,Lesgourgues:2002hk,Hu:2000ke}.
An ultra-light scalar field is motivated as DM because it can
alleviate some of the problems that arise at galactic scale in the
standard paradigm of cold DM, namely, the origin of cuspy
halos~\cite{Moore:1999nt} and the overproduction of
substructure~\cite{Lin:2000qq}. 

Neutrino interaction with DM, $\nu$-DM for short, could have strong
implications at cosmological scales. Interactions of neutrinos with light scalar
fields have been studied with some interesting implications noticed,
such as a reduction of the relic neutrino density, leading to a
neutrinoless universe~\cite{Beacom:2004yd}, or a modification on the
CMB spectra~\cite{Mangano:2006mp,Serra:2009uu}, or even a
connection between the smallness of neutrino mass and a MeV-mass scalar
field DM~\cite{Boehm:2006mi}.
Furthermore, $\nu$-DM interaction might affect the flux of UHE neutrinos.
In particular, such interaction may suppress the neutrino flux
resulting in a kind of GZK cutoff for neutrinos.
Many DM candidates were analyzed
in this context: heavy neutrinos as dark
matter~\cite{Weiler:1983xx,Weiler:1992fm,Roulet:1992pz}, lightest
supersymmetric particles (LSP) discussed in Ref.~\cite{Weiler:1992fm}
and updated in Ref.~\cite{Barenboim:2006dj}, and MeV-mass
scalar field~\cite{Boehm:2004,Mangano:2006mp,Boehm:2006mi}. In all
these cases the suppression is small, not interfering in the
propagation of UHE neutrinos. 

Nevertheless, we show here that a coupling between relic ultra-light
scalar fields and neutrinos may imply a suppression of the UHE
neutrino flux, in which case there may be a confusion between the flux
limit at the source and a reduction of the UHE neutrino flux during
propagation.
Previous analysis have put constraints on the $\nu$-DM interaction
couplings for those models by using, for instance, SN1987A neutrino
data or possible imprints on the angular power spectra of CMB
anisotropies \cite{Mangano:2006mp}. Nevertheless, those limits do not
apply to our case, since they were obtained by assuming a mass of the
scalar field $m_\phi > 10$ MeV while we will explore the possibility
that ultra-light scalar field with $m_\phi \ll 1$ eV can couple to
neutrinos.  Such a small mass for the scalar particle gives a cross
section with different behavior compared the one previously reported
in other works~\cite{Boehm:2004,Boehm:2003hm}, allowing for new
phenomena like a flux suppression for reasonable values of the
coupling constants. Consequences of this type of ultra-light
scalar fields have been studied in other astrophysical contexts like
in the equilibrium of degenerate stars~\cite{Grifols:2005kv}. 

\begin{figure}
\center{
\begin{picture}(80,80)(0,50)
\ArrowLine(0,70)(60,90)
\DashArrowLine(0,120)(60,90){5} 
\Text(5,65)[]{F}
\ArrowLine(60,90)(100,90)
\Text(105,85)[]{f}
\Text(5,110)[]{$\phi$}
\Text(180,90)[]{$g_{\nu\phi} P_R$}
\end{picture}
}
\caption{Feynman rule for the interaction of a neutrino with an ultra-light scalar field.}
\label{feyn-rule}
\end{figure}
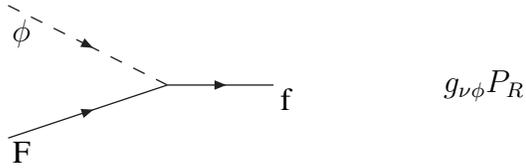

In this work we use the $\nu$-DM coupling as
described by the Feynman rule \cite{Boehm:2003hm} shown in figure~\ref{feyn-rule}
where $\,P_R\,$ denote the chiral projector,
$\phi$ stands for the spin-0 dark matter field
and $g_{\nu\phi}$ denote the strength of its Yukawa coupling. 

In the next Section we show how the neutrino flux may vary due to a
nonstandard interaction of neutrinos with DM. We also introduce the
space of parameters, namely, cross section and DM mass, that lead to
an important effect on the neutrino propagation through cosmological
distances. In Section~\ref{crosssec} we show the elastic scattering
cross section for self and non-self-conjugate scalar fields while
Section~\ref{couplimits} is devoted to study the suppression to the
neutrino flux due to the proposed $\nu$-DM interaction. Finally in
Section~\ref{conclusion} we discuss our results and conclusions.

\section{Neutrino flux and Dark Matter density}
Once the neutrinos are produced at extragalactic sources, they have to
propagate through distances of the order of 100-1000~Mpc to arrive at
the Earth. For this scale of distances it is a good approximation to
consider the distribution of sources and DM as homogeneous and
isotropic. Then, considering $\nu$-DM interaction, one can calculate
the neutrino mean free path $\lambda = (n\sigma)^{-1}$, where $n$ is
the DM particle density ($\rho_{\rm DM}/m_{\rm DM}$) and $\sigma$ is
the $\nu$-DM cross section.  Therefore, at a distance $L$ from the
source the total flux expected is given by:
\begin{equation}\label{fluxfinal}
F(L)=F_0e^{-L/\lambda}\,,
\end{equation}
where $F_0$ is the flux when no interaction with the dark matter medium 
is considered. From Eq.~(\ref{fluxfinal}) we
learn that, e.g., a mean free path of approximately one third of the
mean distance to the sources ($\lambda \sim L/3$) gives a 95\%
suppression of the initial flux. We do not show the result considering
the evolution of sources because it does not change the result
considerably and do not affect our conclusions.

\begin{figure}[t]
\centering
\includegraphics[width=\textwidth]{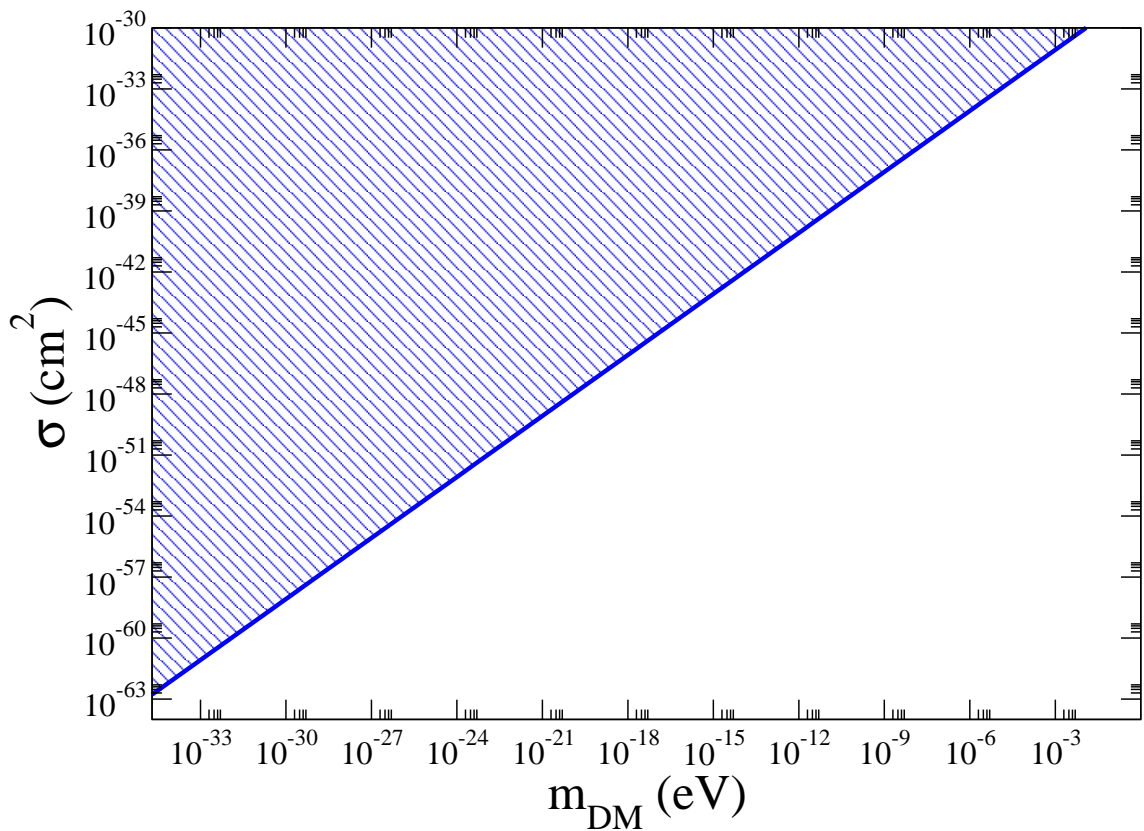}
\caption{Neutrino cross section inducing a 95\% neutrino flux
  suppression, as a function of the DM particle's mass $m_{\rm
  DM}$. The shaded area shows the suppression parameters' space.
  We consider a density $\rho_{\rm DM} = 1.2\times10^{-6}$~GeV/cm$^3$
  and a mean source distance of $L=100$~Mpc. }\label{crossvsDM} 
\end{figure}

Considering, for instance, $\lambda = L/3 \sim 33$~Mpc, we can
compute the cross section $\sigma$ as a function of
the DM particle's mass $m_{\rm DM}$. We consider for the mean DM
density in the universe, $\rho_{\rm DM}=1.2\times10^{-6}$~GeV/cm$^3$~\cite{wmap}.
In figure~\ref{crossvsDM} we show the space of parameters that results
in a 95\% neutrino flux suppression or more.

We observe that there is a flux suppression even for a very small
cross section provided that $m_{\rm DM}$ is also very small.  So we
may conclude that even very weakly interacting particles may have an
effect on the neutrino flux if the mass of the background DM particle
is extremely small, giving a high DM number density $n$.

Such an ultra-light particle has already been considered.  DM scalar
field candidates with mass in the range $10^{-22}-10^{-24}$~eV have
been proposed as viable DM
particles~\cite{Matos:2008ag,Arbey:2001qi,Amendola:2005ad,sflocal,Lesgourgues:2002hk,Hu:2000ke}. Even
lighter scalar fields, with masses lower than $10^{-33}$~eV, have been
postulated in order to explain
DE~\cite{Weiss:1987xa,Frieman:1995pm,Kim:1998kx,Hall:2005xb}.

Concerning ultra-light scalar fields as DM, they have been
studied for a number of self-interaction potentials, like quadratic
ones~\cite{Matos:2008ag}.  The main idea behind these models is that
scalar fields were unified fields at a very early time after the origin of
the universe.  As the universe expands, the scalar fields cool
together with the rest of the particles and finally they decouple
from the rest of the matter.  They condensate once they reach a
critical temperature $T_C$.  For the case of a complex scalar field
the critical temperature is given by~\cite{UrenaLopez:2008zh}
\begin{equation}
T_C=\sqrt{\frac{3q}{m_\phi}}\,,
\end{equation}
where $q$ is the charge density, defined as the excess of particles
$n$ over antiparticles $\bar n$, $q=n-\bar n$, and $m_\phi$ is the mass
of the scalar field. From this formula one sees that an asymmetry
of scalar particles, $n$, over antiparticles, $\bar n$, is required in
order to have a high $T_C$. One can make an
estimate~\cite{Lundgren:2010sp} for the value of $T_C$ considering
$n\gg \bar n$, in which case the antiparticle
contribution to the dark matter density is negligible and, therefore,
\begin{equation}
\rho_{\rm DM} \simeq n m_\phi\,.
\end{equation}

If we consider that the present dark matter density is $\rho_{\rm
DM} \simeq 0.23 \rho_c$ and $\rho_c \simeq 4.19 \times 10^{-11}~{\rm
eV}^4$ we conclude that $n \simeq 10^{12}~{\rm eV}^3$. For a
scalar field mass of $m_\phi = 10^{-23}~{\rm eV}$ we get a critical
temperature of condensation $T_c \simeq 10^{17}~{\rm eV}$.

In this example, condensation occurs at very early stages of the
evolution of the universe. After the scalar field condenses, most of
these bosons lie in the ground state and one coherent field is
appropriate to describe its evolution as the universe expands.  It
was shown in~\cite{Turner:1983he} that, for a coherent scalar field
with a potential $V(\phi) \simeq \phi^k$, the energy density decreases
as $\rho_\phi \sim a^{-6k/(k+2)}$, with $a$  the cosmic scale factor.  For
our case of interest of scalar field DM~\cite{Matos:2008ag}, the potential is
$V(\phi) \sim \phi^2$ and then, the energy density decreases as
$\rho_{\rm \phi} \sim a^{-3}$, i.e., it evolves as dust and hence as
cold dark matter.

We show our results for two different situations: when the total
amount of DM in the universe is composed of the ultra-light scalar
field $(\rho_\phi=\rho_{\rm DM})$ and for the case of a
multi-component DM in the universe~\cite{Amendola:2005ad}, where the
ultra-light scalar field density is a ten percent fraction of the
total density in the $\Lambda$CDM model.  Therefore, in this case
\begin{equation}\label{eq:density}
n=0.1\frac{\rho_{\rm DM}}{m_\phi}\,,
\end{equation}
where $m_\phi$ is the scalar field mass and the ultra-light DM density
is given by $\rho_\phi=0.1\rho_{\rm DM}$.
Any other DM fraction can be obtained by simply rescaling the results.

\section{The cross section}\label{crosssec}
The kind of interaction we are assuming is the elastic scattering $\nu
+\phi \to \nu+\phi$,  where $\phi$ is the ultra-light scalar field.
To compute an expression for the cross section
in this process, we assume one of the models proposed in
ref.~\cite{Boehm:2003hm},  where the DM candidate can be either
self-conjugate ($\phi=\phi^*$) or non-self-conjugate
($\phi\ne\phi^*$).

The Lagrangian for the interaction is given by
\begin{equation}\label{eq:lagrange}
{\cal L}=g_{\nu\phi}\,\bar\nu\,\phi\,P_RF + H.c.\,,
\end{equation}
where $g_{\nu\phi}$ is the $\nu$-DM coupling, $P_R$ denotes the chiral
projector $(1+\gamma^5)/2$ and $F$ denotes a new spin one half fermion
that mediates the interaction.

\subsection{Non-self-conjugate scalar field dark matter ($\phi \ne \phi^*$)}
For this case, only the $u$-channel contributes to the cross section
amplitude~\cite{Boehm:2003hm}, and it is given by:
\begin{equation}\label{elasticCS}
\sigma \simeq \frac{g_{\nu\phi}^4}{32\pi}\frac{s}{(u-M_I^2)^2}\,,
\end{equation}
where the center of mass energy is $\sqrt s=\sqrt{2 m_\phi E_\nu}\,$,
$E_\nu$ is the neutrino energy, and $M_I$ is the mass of the
intermediate particle for the $\nu$-DM interaction. Since we are
considering an ultra-light scalar field, $u << M_I^2$ for all the
considered energy range.  In that limit the cross section can be
written in the form:

\begin{equation}\label{sfield-cs}
\sigma \simeq \left(\frac{g_{\nu \phi}}{M_I}\right)^4 \frac{m_\phi E_\nu}{16 \pi} \,.
\end{equation}
\subsection{Self-conjugate scalar field dark matter ($\phi=\phi^*$)}
In this case there is a contribution from the $s$-channel
\cite{Boehm:2003hm}.  Neglecting the neutrino mass, in the local limit
approximation ($u,s \ll M_I^2$), both contributions from $u$ and $s$
channel cancel each other.  But considering the neutrino mass the
cross section, although small, is not exactly zero and it is given by the
expression:

\begin{eqnarray}\label{CSauto}
\frac{d\sigma}{d\Omega}&=&\frac{g_{\nu\phi}^4}{32\pi^2} [
  \frac{m_\nu^2}{4} \left(\frac{1-\cos
    \theta}{(s-M_I^2)^2}+\frac{1-\cos \theta}{(u-M_I^2)^2}\right)
  \nonumber \\ &+&\left(\frac{1}{s-M_I^2}-\frac{1}{u-M_I^2}
  \right)^2 \left(\frac{s}{4}(1+\cos \theta)-\frac{m_\phi^2}{4}(1-\cos
  \theta)\right) \nonumber
  \\ &+&\left(\frac{1}{s-M_I^2}-\frac{1}{u-M_I^2}\right)m_\nu^2\left(\frac{1+\cos
    \theta}{2(s-M_I^2)} +\frac{1}{u-M_I^2}\right) \nonumber
  \\ &+&\frac{2m_\nu^4}{s(u-M_I^2)(s-M_I^2)} ]\,.
\end{eqnarray}

In the limit where $s,u \ll M_I$ the cross section, after integration
in solid angle, is reduced to

\begin{equation}\label{selfconjugatedCS}
\sigma\simeq\left(\frac{g_{\nu\phi}}{M_I}\right)^4\frac{m_\nu^2}{16 \pi}\,.
\end{equation}

In our computation of Eqs.~(\ref{elasticCS}) and (\ref{CSauto}) we
consider, for simplicity, the limit $m_\nu^2 << s$, which for a
neutrino energy of the order of $E_\nu\sim 10^{18}$~eV, and neutrino
mass around $1$~eV, translates into a restriction for the mass of the
scalar field DM $m_\phi >> {\cal O}~(10^{-18})$~eV.  We remark that,
however, it is possible to consider smaller masses for the scalar
field DM, recalculating the cross section without working in this
limit.

\section{Neutrino flux suppression}\label{couplimits}
\subsection{Non-self-conjugate scalar field dark matter}\label{subnudmlim}

Remembering that $\lambda=(n\sigma)^{-1}$ and using
Eq.~(\ref{sfield-cs}) for $\sigma$, we have the neutrino mean free
path given by
\begin{eqnarray}\label{eq:mfp2}
\lambda&=&16 \pi \left(\frac{M_I/g_{\nu \phi}}{\rm GeV}\right)^4\left(
\frac{\rm GeV}{E_\nu}\right)\left(\frac{{\rm GeV/cm}^3}{\rho_\phi} \right)
{\rm GeV}^2{\rm cm}^3 \nonumber \\
&\simeq& L_0
\left(\frac{M_I/g_{\nu \phi}}{\rm GeV}\right)^4\left(
\frac{10^{18}{\rm eV}}{E_\nu}\right)\left(\frac{{\rm GeV/cm}^3}{\rho_\phi} \right)\,,
\end{eqnarray}
where $L_0\simeq42$~pc.

Depending specially on the $\nu$-DM coupling strength, a neutrino flux
suppression is possible.  In this case, the non observation of UHE
neutrino events by the experiments do not mean that there is a limit
for the neutrino production in the extragalactic sources, but rather a
reduction of the neutrino flux due to the $\nu$-DM interaction. We can
estimate the necessary strength of the coupling $g_{\nu \phi}$ to
produce an important effect on the neutrino flux for propagation
through distances of the order of 100~Mpc or more. It is also
important to notice that, in this case, the mean free path is
independent of the mass of the scalar field DM.

From Eqs.~(\ref{fluxfinal}) and~(\ref{eq:mfp2}) we conclude that if
the interaction have at least a strength given by the coupling
\begin{equation}\label{eq:coupling}
\frac{g_{\nu \phi}}{M_I} \gtrsim \left[
\ln\left(\frac{F_0}{F}\right)\frac{L_0}{\rho_\phi E_\nu L}
\right]^\frac{1}{4}\,,
\end{equation}
the source flux limit is loose due to the above argument. In
Eq.~(\ref{eq:coupling}) $L$ is given in Mpc, $E_\nu$ in GeV,
$\rho_\phi$ in GeV/cm$^3$, and $M_I$ in GeV.

\begin{figure}[t]
\centering
\includegraphics[width=\textwidth]{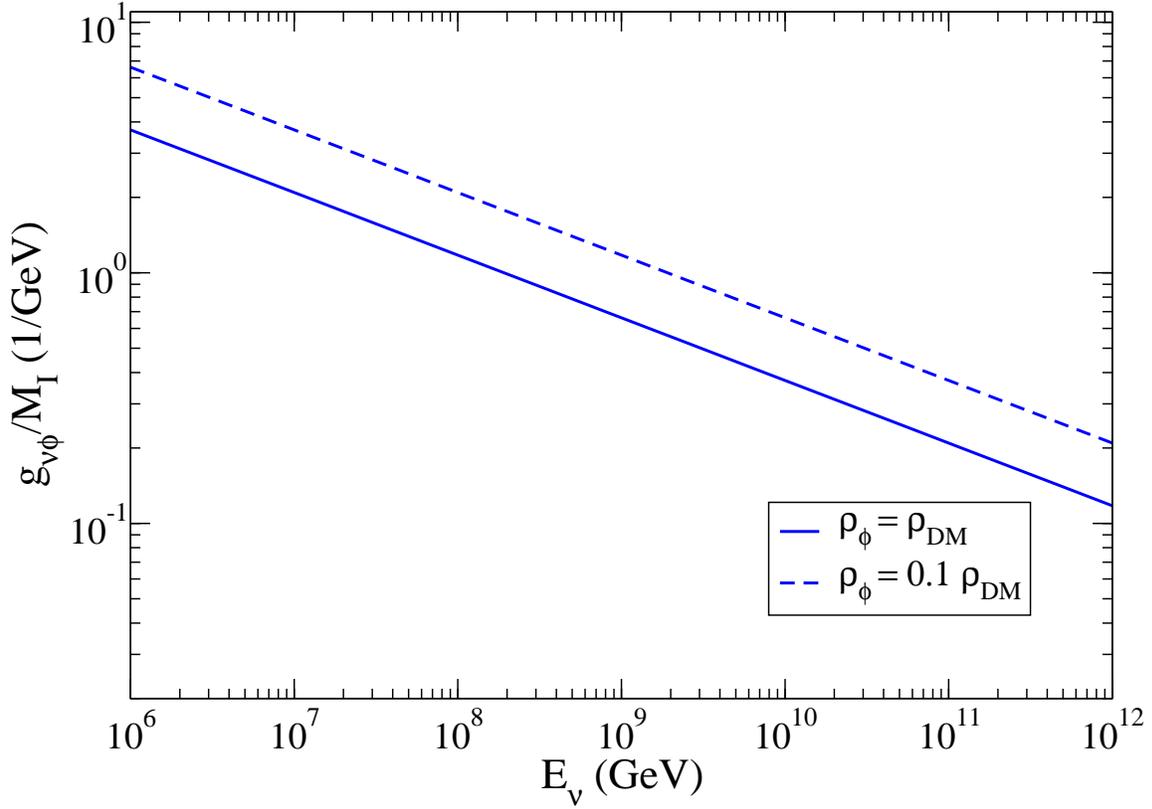}
\caption{Ratio $\frac{g_{\nu \phi}}{M_I}$ as a function of the
  neutrino energy $E_\nu$ that induces a neutrino flux suppression,
  for the non-self-conjugate dark matter forming the total amount
    of DM (solid line) or a 10\% fraction (dashed line).  The regions
    above the curves predict stronger suppression.  The considered
  mean distance to the sources is $L=5\times 10^2$~Mpc and the
  neutrino flux with no suppression is taken according to the MPR
  limit prediction.}\label{suppression}
\end{figure}

We show this result graphically in
figure~\ref{suppression}. Considering a non suppressed neutrino flux
$F_0$, equal to the one calculated by Mannheim, Protheroe, and Rachen
(MPR)~\cite{Mannheim:1998wp}, we show in this figure the value of the
ratio $\frac{g_{\nu \phi}}{M_I}$ that leads to a suppression giving a
neutrino flux, $F$, equal to the Waxman and Bahcall (WB)
bound~\cite{Waxman:1998yy,Bahcall:1999yr}. The solid line is for a
single ultra-light field forming the DM in the universe. The dashed
one is for a 10\% ultra-light DM component.

According to the WB bound, the maximum allowed neutrino flux is of the
order of ${\cal
  O}(10^{-8})\varepsilon_Z$~GeV\,cm$^{-2}$s$^{-1}$sr$^{-1}$, where
$\varepsilon_Z$ is of order unity and includes possible contribution
of so far unobserved high redshift sources and the effect of redshift
in neutrino energy.
In this model protons are confined in the astrophysical sources and 
undergo photoproduction of mesons and neutrons. The mesons' decay 
generates the neutrino flux,  while the neutrons escape from the
acceleration site, decay, and produce the observed UHE cosmic rays.

On the other hand, the MPR flux model consider neutron optically thick
sources ($\tau_{n\gamma}>>1$), i.e., in the photoproduction process
neutrinos escape from the sources, but not the cosmic rays. Therefore,
the relation between the neutrino and the cosmic ray fluxes is not
direct and the neutrino flux limit is relaxed up to ${\cal O}
(10^{-6})$~GeV\,cm$^{-2}$s$^{-1}$sr$^{-1}$.
For these models it corresponds to a total neutrino flux ratio at the
Earth of the order of $F_{\rm WB}/F_{\rm MPR}\approx{\cal O}(0.01)$,
which is the ratio we use in Eqs.~(\ref{eq:coupling}) and
(\ref{eq:couplingself}). For a recent discussion of the neutrino flux
limits from various experiments see~\cite{Brunner:2011ji}.

The value of $L$ is fixed to be $L=5\times 10^2$~Mpc. One can see that
an ultra-light scalar field, if it exists as DM in the universe, may
generate an extragalactic neutrino flux suppression effect. For
instance, for a 10\% scalar field DM component, if the ratio
$\frac{g_{\nu \phi}}{M_I}$ is of the order 0.1, neutrinos with
energies of the order of $10^{18}$~eV would present a suppression
reducing the flux from the MPR to the WB value.

\subsection{Self-conjugate scalar field dark matter}
A similar analysis can be made for the cross section given in Eq.
(\ref{selfconjugatedCS}), valid for the case of self-conjugate scalar
field dark matter. The resulting neutrino mean free path is

\begin{eqnarray}\label{eq:mfpself}
\lambda&=&16 \pi \times 10^{-6}\left(\frac{M_I/g_{\nu \phi}}{\rm GeV}\right)^4\left(
\frac{\rm eV}{m_\nu}\right)^2\left(\frac{{\rm GeV/cm}^3}{\rho_\phi} \right)
\left(\frac{m_\phi}{10^{-15}\rm eV}\right)
{\rm GeV}^2{\rm cm}^3 \nonumber \\
&\simeq& L_0\left(\frac{M_I/g_{\nu \phi}}{\rm GeV}\right)^4\left(
\frac{\rm eV}{m_\nu}\right)^2\left(\frac{{\rm GeV/cm}^3}{\rho_\phi} \right)
\left(\frac{m_\phi}{10^{-18}\rm eV}\right)\,,
\end{eqnarray} 
where $L_0\simeq42$~pc. 
The required ratio $\frac{g_{\nu \phi}}{M_I}$
in order to have a reduction of the flux $F_0$, with no suppression mechanism, 
to a flux $F$, or lower, due to the $\nu$-DM interaction should be

\begin{equation}\label{eq:couplingself}
\frac{g_{\nu \phi}}{M_I} \gtrsim \left[
\ln\left(\frac{F_0}{F}\right)\frac{L_0 m_\phi}{\rho_\phi m_\nu^2 L}
\right]^\frac{1}{4}\,.
\end{equation} 

\begin{figure}[t]
\centering
\includegraphics[width=\textwidth]{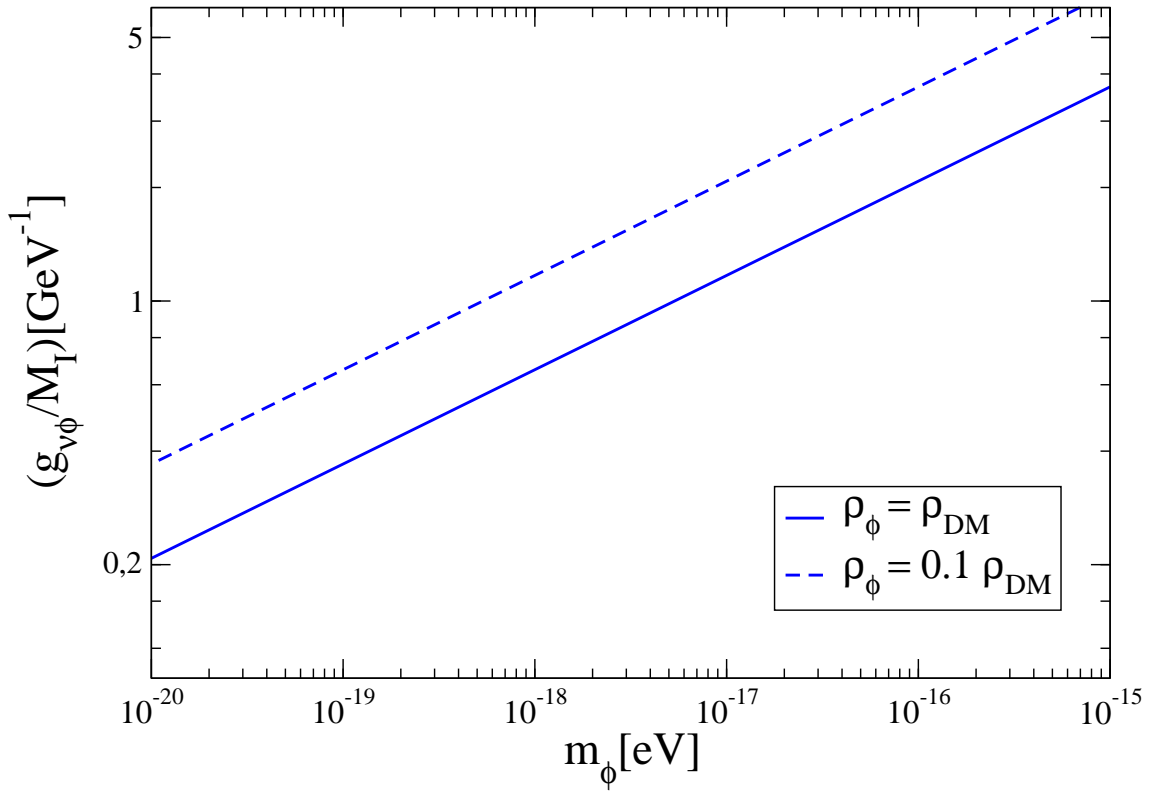}
\caption{Ratio $\frac{g_{\nu \phi}}{M_I}$ as a function of the mass
of the scalar field DM $m_\phi$ that induces a neutrino flux suppression
(regions above the curves), for the self-conjugate dark matter
forming the total amount of DM (solid line) or a 10\% fraction (dashed line). 
We consider the mean distance to the sources 
$L=5\times 10^2$~Mpc, a neutrino mass of
$1$~eV, and the neutrino flux with no suppression is taken
according to the MPR limit prediction.}\label{Fig3}
\end{figure}

The above ratio is shown in figure \ref{Fig3} as a function of
the mass of the scalar field DM candidate, using $L=5\times 10^2$~Mpc
and $m_\nu=1$~eV.  We assume again $F_0$ equal to the 
MPR prediction, while the suppressed flux, $F$, is considered to be 
equal or below the WB limit. The solid line
is for a single ultra-light field forming the DM in the universe. The dashed one
is for a 10\% ultra-light DM component.

\section{Discussion and Conclusion}\label{conclusion}
There are several experiments, like IceCube and the Pierre Auger
Observatory, expecting to detect extragalactic neutrinos.  But
neutrinos with energies above $10^{15}$~eV, coming from extragalactic
sources, have not been detected yet.  We study the possibility that
UHE neutrinos could be absorbed while traveling from their sources to
the Earth. In particular we illustrate this idea by considering an
ultra-light particle as a component of the Dark Matter in
the Universe.

We consider a mechanism for the neutrino interaction based on a scalar
field dark matter model and we show that in this case the propagation
of extragalactic neutrinos from sources $100$~Mpc or farther from the
Earth may be affected. This would give negative results on neutrino
telescopes or UHE neutrino detectors. On the other hand, 
despite neutrinos from a nearby supernova could interact
with the DM halo around the collapsing star, the scale involved
would not be sufficient for an absorption like the one proposed here.

Although nonstandard interactions of neutrinos with
Dark Matter particles had been considered in the literature before,
no important effect on neutrino propagation had been predicted.
In most of the literature, light scalar field DM had been
considered to be relativistic and the coupling to neutrinos was
constrained due to measurable effects on the CMB spectra.  In this
work we have considered non-relativistic ultra-light scalar fields,
proposed in the literature, that besides their gravitational
effects, may not have other measurable astrophysical
consequences.  

To our knowledge, this is the first example of a possible suppression
of the extragalactic neutrino flux due to propagation effects.
Therefore, care must be taken when
using the limits obtained by such experiments, since those limits can
be due to two factors: source limit and/or absorption due to
UHE neutrino-ultra light scalar field Dark Matter interaction during
neutrino propagation from the source to the Earth. On the other hand,
a positive signal of UHE neutrinos could be useful to put restrictions
on models that contains a light scalar field DM candidate.

Similar arguments could be applied for particles other than
neutrinos. For instance, in~\cite{Andrianov:2009kj} it was analyzed the
suppression of charged particles due to the interaction with a
pseudoscalar and it was shown that the axion can play the
role of a shield for high energy cosmic rays.

\acknowledgments
  This work has been supported by CONACyT grant 132197, SNI-Mexico and
  PAPIIT grant IN117611-3.  T.~I.~Rashba and C.~A.~Moura thank the
  Physics Department of CINVESTAV for the hospitality during the visit
  when part of this work was done. F.~Rossi-Torres thanks CAPES and
  CNPq for the financial support.

\end{document}